\documentclass[prd,aps,showpacs,preprintnumbers,amssymb]{revtex4}
\usepackage{graphicx}% Include figure files
\usepackage{epsf}
\usepackage{amsmath}
\usepackage{epstopdf}
\usepackage{multirow}
\usepackage{bm}
\usepackage{color}
\def\e3p{$\eta \rightarrow 3 \pi$}

\begin{document}
\title{%
\hfill{\normalsize\vbox{%
\hbox{}
 }}\\
{Trace and axial anomalies on equal footing }}

\author{Renata Jora
$^{\it \bf a}$~\footnote[1]{Email:
 rjora@theory.nipne.ro}}

\affiliation{$^{\bf \it a}$ National Institute of Physics and Nuclear Engineering PO Box MG-6, Bucharest-Magurele, Romania}

\date{\today}

\begin{abstract}
We discuss for some particular non supersymmetric theories a generalized symmetry that includes both the scale and axial transformations and  leads to a single current that may contain also a pseudoscalar term. The method, inspired by the superconformal anomalies  has important application for low energy effective models where it allows the introduction of  a single complex glueball field with a scalar and a pseudoscalar component  on the same footing with the complex meson nonets fields made of quarks. Both axial and trace and axial anomalies are satisfied in accordance to the meson structure and the QCD requirements.

\end{abstract}
\maketitle

\section{Introduction}

One of the most important tools in dealing with physical systems and especially quantum field theoretical models is the incorporation of symmetries, global or gauged, exact or approximate.   We know that a standard Lagrangian must be real and thus hermitian and Lorentz invariant. In some instances the Lagrangian might be also scale invariant at least at tree level thus leading to new class of theories whose properties were explored in detail in the literature. The scale invariance is usually broken at the quantum level \cite{Coleman}-\cite{Collins} by the renormalization group equations  and at tree level by explicit noninvariant terms in the Lagrangian. One usually introduces the symmetric energy momentum tensor $\theta^{\mu}_{\nu}$  which may be derived with some complications or not from the canonical energy momentum tensor $T^{\mu}_{\nu}$. Then the amount of breaking of the scale invariance is measured by:
\begin{eqnarray}
\partial_{\mu}D^{\mu}=\theta^{\mu}_{\mu},
\label{scaleinfttr54663}
\end{eqnarray}
where $D^{\mu}=\theta^{\mu}_{\nu}x^{\nu}$ is the dilatation current and $\partial_{\mu}\theta^{\mu}_{\nu}=0$.

The quantum anomalies of a theoretical particle physics model are crucial for describing its properties and more so for the supersymmetric  QCD and the QCD Lagrangians where they are an important tool for constructing low energy effective models of hadrons.

The effective approach based on the exact realization of axial and scale anomalies for supersymmetric QCD was analyzed by Seiberg \cite{Seiberg1} and Seiberg and Witten \cite{Seiberg2} leading to remarkable results regarding the phase structure of these theories. Application of these methods for QCD were then implemented in \cite{Schechter1}, \cite{Schechter2}.

It is known that in supersymmetric gauge theories at least in the holomorphic picture the axial and trace anomalies belong to the same supermultiplet and thus the pseudoscalar and scalar glueball can be regarded on the same footing.  A unified description of both these anomalies for supersymmetric gauge theories was presented in \cite{Rose}. For example for supersymmetric  Yang Mills ($F^a_{\mu\nu}$ is the gauge field tensor),
\begin{eqnarray}
&&\theta_{\mu}^{\mu}=3N_c(F+F*)=-\frac{3N_cg^2}{32\pi^2}F^{a\mu\nu}F^a_{\mu\nu}
\nonumber\\
&&\partial^{\mu}J_{\mu}^5=2iN_c(F-F^*)=\frac{N_cg^2}{32\pi^2}F^{a\mu\nu}\tilde{F}^a_{\mu\nu}.
\label{supersyman65774}
\end{eqnarray}
Here $J_{\mu}^5=\bar{\lambda}\bar{\sigma}_{\mu}\lambda_a$ is the axial anomaly current and $\lambda$ is the gluino field. Moreover the main ingredient of the effective theory is $S$, a composite chiral superfield with the structure:
\begin{eqnarray}
S(y)=\Phi(y)+\sqrt{2}\theta\Psi(y)+\theta^2F(y),
\label{chiralsuerf7775}
\end{eqnarray}
where,
\begin{eqnarray}
&&\Phi\approx \lambda^2
\nonumber\\
&&\Psi\approx \sigma^{\mu\nu}\lambda_aF_{\mu\nu}^a
\nonumber\\
&&F\approx-\frac{1}{2}F^{a\mu\nu}F^a_{\mu\nu}-\frac{i}{4}\epsilon_{\mu\nu\rho\sigma}F^{a\mu\nu}F^{a\rho\sigma}.
\label{def4553}
\end{eqnarray}

Inspired by supersymmetric QCD  we will present here an approach for QCD where  the pseudoscalar and scalar glueballs are treated as components of a single complex glueball field leaving for future work any possible direct connection with supersymmetry and the subtleties that might arise from it.

In section II we discuss from a new point of view the scale  symmetry and the construction of the symmetric energy momentum tensor for real and complex scalar field theories and QED. In section III we will slightly modify the scale transformation such that to incorporate also the axial transformation for QED \cite{Adler2}-\cite{Bell} and the $U(1)$ transformation for the complex scalar field theory. Thus we will add to the symmetric energy momentum tensor presumably a "pseudoscalar" tensor $\tau^{\mu}_{\nu}$  such that the associated full Noether current becomes:
\begin{eqnarray}
J^{\mu}=\theta^{\mu}_{\nu}x^{\nu}+\tau^{\mu}_{\nu}x^{\nu},
\label{toatlcurrent54663}
\end{eqnarray}
with:
\begin{eqnarray}
&&\partial_{\mu}\theta^{\mu}_{\nu}=0
\nonumber\\
&&\partial_{\mu}\tau^{\mu}_{\nu}=0.
\label{trsn4663}
\end{eqnarray}
Finally the conservation of currents for the extended transformation will be:
\begin{eqnarray}
\partial_{\mu}J^{\mu}=\theta^{\mu}_{\mu}+\tau^{\mu}_{\mu}.
\label{cons66488}
\end{eqnarray}

In section IV we will show how this approach makes total sense for a low energy QCD linear sigma model with two chiral nonets, model discussed in detail in \cite{Jora1}-\cite{Jora8}. Section V is dedicated to Conclusions.

\section{Scale transformation revisited}

We will analyze in some details the scale transformation for some simple theories with the amend that our findings can be generalized easily to more intricate models.

\subsection{Scalar field theory}

We start with  a Lagrangian for the scalar field theory:
\begin{eqnarray}
{\cal L}=\frac{1}{2}\partial^{\mu}\Phi\partial_{\mu}\Phi-V(\Phi).
\label{lagrsc54663}
\end{eqnarray}
Under the scale transformation the scalar field transforms as:
\begin{eqnarray}
\Phi(x)\rightarrow \exp[-\sigma]\Phi(x\exp[-\sigma]),
\label{scfitr663554}
\end{eqnarray}
whereas the action changes to:
\begin{eqnarray}
&&\delta S=
\int \delta(d^4x){\cal L}(\Phi(x),x)+
\nonumber\\
&&\int d^4x \frac{\partial{\cal L}}{\partial x^{\mu}}\delta x^{\mu}+\int d^4x \partial_{\mu}\Bigg[\frac{\partial {\cal L}}{\partial \partial^{\mu}\Phi(x)}\delta(\Phi(x))\Bigg].
\label{dretytuu5674}
\end{eqnarray}
The conserved current can be determined to be:
\begin{eqnarray}
J_{\mu}=\frac{\partial {\cal L}}{\partial \partial_{\mu}\Phi(x)}\Phi(x)+\Bigg[\frac{\partial {\cal L}}{\partial \partial_{\mu}\Phi(x)}\partial_{\rho}\Phi(x)-{\cal L}\eta_{\rho}^{\mu}\Bigg] x^{\rho}.
\label{currentttsc54663}
\end{eqnarray}
The quantity,
\begin{eqnarray}
\frac{\partial {\cal L}}{\partial (\partial_{\mu}\Phi(x))}\partial_{\rho}\Phi(x)-{\cal L}\eta^{\mu}_{\rho}=T^{\mu}_{\rho},
\label{deftyy667}
\end{eqnarray}
is the canonical energy momentum tensor.  There are different procedures for deriving the symmetric energy momentum tensor $\theta^{\mu}_{\rho}$ out from the canonical energy momentum tensor in Eq. (\ref{deftyy667}). Here we will present a very simple way applicable to any theory in general. First we observe that we need to add an extra term to it that can stem only from the first term in the expression (\ref{currentttsc54663}). We write:
\begin{eqnarray}
\frac{\partial {\cal L}}{\partial(\partial_{\mu}\Phi)}\Phi=a\partial^{\mu}\Phi\Phi\partial_{\nu}x^{\nu}-a\partial_{\rho}\Phi\Phi\eta^{\mu}_{\nu}\partial_{\rho}x^{\nu}=3a\partial^{\mu}\Phi\Phi.
\label{exptr553454}
\end{eqnarray}
Since Eq. (\ref{exptr553454}) is an identity we immediately determine $a=\frac{1}{3}$. Then one can write:
\begin{eqnarray}
&&\frac{\partial {\cal L}}{\partial(\partial_{\mu}\Phi)}\Phi=
\frac{1}{3}\partial^{\mu}\Phi\Phi\partial_{\nu}x^{\nu}-\frac{1}{3}(\partial^{\rho}\Phi)\Phi\eta^{\mu}_{\nu}\partial_{\rho}x^{\nu}=
\nonumber\\
&&\frac{1}{3}\partial_{\nu}[\partial^{\mu}\Phi\Phi x^{\nu}]-\frac{1}{3}\partial_{\rho}[(\partial^{\rho}\Phi)\Phi\eta^{\mu}_{\nu}x^{\nu}]-
\nonumber\\
&&\frac{1}{3}\partial_{\nu}[(\partial^{\mu}\Phi)\Phi] x^{\nu}+\frac{1}{3}\partial_{\rho}[(\partial^{\rho}\Phi)\Phi\eta^{\mu}_{\nu}]x^{\nu}.
\label{resul64773}
\end{eqnarray}
We drop the total derivatives from the current to get:
\begin{eqnarray}
\frac{\partial {\cal L}}{\partial(\partial_{\mu}\Phi)}\Phi=\Bigg[-\frac{1}{6}\partial_{\nu}\partial^{\mu}(\Phi^2)+\frac{1}{6}\partial^2\Phi^2\eta^{\mu}_{\nu}\Bigg]x^{\nu}.
\label{frn63299}
\end{eqnarray}
Then the symmetric energy momentum tensor is just:
\begin{eqnarray}
\theta^{\mu}_{\nu}=T^{\mu}_{\nu}+\Bigg[-\frac{1}{6}\partial_{\nu}\partial^{\mu}(\Phi^2)+\frac{1}{6}\partial^2\Phi^2\eta^{\mu}_{\nu}\Bigg],
\label{symet54663}
\end{eqnarray}
and this result is consistent with the standard formula for the symmetric energy momentum  tensor in the literature.

Next we consider a complex scalar field theory with the Lagrangian:
\begin{eqnarray}
{\cal L}=\partial^{\mu}\Phi^*\partial_{\mu}\Phi-V(\Phi,\Phi^*).
\label{complex637884}
\end{eqnarray}
We consider an extension of the scale transformation in which $x'=\exp[\sigma] x$ and the fields transform as:
\begin{eqnarray}
&&\Phi(x)\rightarrow\Phi'(x')=\exp[-\sigma(1+i)]\Phi(\exp[-\sigma] x)
\nonumber\\
&&\Phi^*(x)\rightarrow\Phi^{\prime *}(x') =
\exp[-\sigma(1-i)]\Phi^*(\exp[-\sigma]x).
\label{res553662}
\end{eqnarray}

Note that the above transformation is just the scale transformation associated with a $U(1)$ transformation with the same infinitesimal parameter $\sigma$. Then it can be easily determined:
\begin{eqnarray}
&&J^{\mu}=\frac{\partial {\cal L}}{\partial (\partial_{\mu}\Phi)}\Phi+\frac{\partial {\cal L}}{\partial (\partial_{\mu}\Phi^*)}\Phi^*+
\nonumber\\
&&\Bigg[\frac{\partial {\cal L}}{\partial (\partial_{\mu}\Phi)}\partial_{\nu}\Phi+\frac{\partial {\cal L}}{\partial (\partial_{\mu}\Phi^*)}\partial_{\nu}\Phi^*-{\cal L}\eta^{\mu}_{\nu}\Bigg]x^{\nu}+
\nonumber\\
&&i\frac{\partial {\cal L}}{\partial (\partial_{\mu}\Phi)}\Phi-i\frac{\partial {\cal L}}{\partial (\partial_{\mu}\Phi^*)}\Phi^*.
\label{compltransfr}
\end{eqnarray}

Finally one can apply the procedure in Eq. (\ref{symet54663}) to obtain:
\begin{eqnarray}
J^{\mu}=\theta^{\mu}_{\nu}x^{\nu}-K^{\mu},
\label{calresuuu56774}
\end{eqnarray}
where,
\begin{eqnarray}
K^{\mu}=-i(\partial^{\mu}\Phi^*)\Phi+i(\partial^{\mu}\Phi)\Phi^*.
\label{def67}
\end{eqnarray}
Our aim is to generalize the trace anomaly such that to contain also a contribution from the term in Eq. (\ref{def67}). Specifically we want to find $\tau^{\mu}_{\nu}$ such that $\partial_{\mu}\tau^{\mu}_{\nu}=0$ and:
\begin{eqnarray}
K^{\mu}=-\tau^{\mu}_{\nu}x^{\nu}.
\label{def54637}
\end{eqnarray}

We find using the procedure in Eq. (\ref{resul64773}):
\begin{eqnarray}
\tau^{\mu}_{\nu}=\Bigg[-\frac{i}{3}\partial_{\nu}(\partial^{\mu}\Phi^*\Phi)+\frac{i}{3}\partial^{\rho}(\partial_{\rho}\Phi^*\Phi)\eta^{\mu}_{\nu}\Bigg]+h.c.
\label{finres663554}
\end{eqnarray}
It can be easily checked that:
\begin{eqnarray}
&&K^{\mu}=-\tau^{\mu}_{\nu}x^{\nu}
\nonumber\\
&&\partial_{\mu}\tau^{\mu}_{\nu}=0.
\label{res66277}
\end{eqnarray}

Finally one can compute the trace of the tensor in Eq. (\ref{res66277}) as:
\begin{eqnarray}
-\partial_{\mu}K^{\mu}=\tau^{\mu}_{\mu}=-i\Bigg[\frac{\partial V}{\partial\Phi}\Phi-\frac{\partial V}{\partial \Phi^*}\Phi^*\Bigg],
\label{finalet6647}
\end{eqnarray}
where we applied the equation of motion.

\subsection{QED}

Now we shall state the results for $\tau^{\mu}_{\nu}$ briefly for QED \cite{Adler2}-\cite{Bell} (for a detailed review see \cite{Peskin}). Consider the transformation of the fields:
\begin{eqnarray}
&&\Psi_L(x)\rightarrow \exp[-\sigma(3/2+i)]\Psi_L(\exp[-\sigma] x')
\nonumber\\
&&\Psi_R(x)\rightarrow\exp[-\sigma(3/2-i)]\Psi_R(\exp[-\sigma] x'),
\label{trsnferm3777466}
\end{eqnarray}
together with the corresponding transformation for the conjugate fields. Since the scale transformation contribution in QED is well known we will consider only the extra contribution coming form the axial tarnsformation included in Eq. (\ref{trsnferm3777466}). Thus the conserved current has the form:
\begin{eqnarray}
&&J_L^{\mu\prime}=J_L^{\mu}-i\frac{\partial {\cal L}}{\partial (\partial_{\mu}\Psi_L)}\Psi_L
\nonumber\\
&&J_R^{\mu\prime}=J_R^{\mu}+i\frac{\partial {\cal L}}{\partial (\partial_{\mu}\Psi_R)}\Psi_R.
\label{res773664}
\end{eqnarray}

Here $J_L^{\mu}$ and $J_R^{\mu}$ are the standard scale transformation currents. We denote,
\begin{eqnarray}
&&K^{\mu}=i\frac{\partial {\cal L}}{\partial (\partial_{\mu}\Psi_L)}\Psi_L-i\frac{\partial {\cal L}}{\partial (\partial_{\mu}\Psi_R)}\Psi_R=
\nonumber\\
&&\bar{\Psi}\gamma^{\mu}\gamma^5\Psi,
\label{res664}
\end{eqnarray}
where we applied the equation of motion. Thus we obtained the standard axial current. Next we want to determine $\tau^{\mu}_{\nu}$ such that:
\begin{eqnarray}
&&K^{\mu}=-\tau^{\mu}_{\nu}x^{\nu}
\nonumber\\
&&\partial_{\mu}\tau^{\mu}_{\nu}=0.
\label{res663554}
\end{eqnarray}
For that there is only one possible term:
\begin{eqnarray}
\tau^{\mu}_{\nu}=\frac{1}{3}\partial_{\nu}(\bar{\Psi}\gamma^{\mu}\gamma^5\Psi)-\frac{1}{3}\partial_{\rho}(\bar{\Psi}\gamma^{\rho}\gamma^5\Psi)\eta^{\mu}_{\nu},
\label{res52899}
\end{eqnarray}
where its is clear that the second equation in Eq. (\ref{res663554}) is satisfied if one applies the equation of motion in the absence of quark masses. Thus the full current will be:
\begin{eqnarray}
J^{\mu\prime}=\Bigg[\theta^{\mu}_{\nu}+\tau^{\mu}_{\nu}\Bigg]x^{\nu},
\label{res773664}
\end{eqnarray}
and we were able to write a total current  that encapsulate both  scale and axial transformations. Finally the conservation law is:
\begin{eqnarray}
\partial_{\mu}J^{\prime\mu}=\theta^{\mu}_{\mu}+\tau^{\mu}_{\mu},
\label{res6463553}
\end{eqnarray}
and includes both a scalar and a pseudoscalar part.

\section{Application to a generalized linear sigma model}

In this section we will show in detail  how one can introduce in a consistent way from the point of view presented here the scalar and pseudoscalar glueballs in a linear sigma model which is a low energy description of QCD. A The presence and properties of  the scalar and pseudoscalar glueballs in the QCD spectrum were subject of relevant works in the literature. In \cite{Giacosa}, \cite{Rebhan} the trace anomaly in low energy QCD and some features of the scalar glueball were discussed in different frameworks. Furthermore in \cite{Giacosa1}\cite{Rebhan1} and \cite{Pisarski} the axial anomaly and spectrum and decays of the pseudoscalar glueball were analyzed in low energy QCD.

As an application of our method we consider a generalized linear sigma model \cite{Jora1}-\cite{Jora7} with two chiral meson nonets one with a quark antiquark structure the other one with a four quark composition.  We denote this nonets by $M$ and $M'$ where $M=S+i\Phi$ and $M'=S'+i\Phi'$ and $S$ and $S'$ represent the scalar states and $\Phi$ and $\Phi'$ the pseudoscalar ones. The fields
 $M$ and $M'$ transform in the same way under
chiral SU(3) transformations
\begin{eqnarray}
M &\rightarrow& U_L\, M \, U_R^\dagger,\nonumber\\
M' &\rightarrow& U_L\, M' \, U_R^\dagger,
\end{eqnarray}
but transform differently under U(1)$_A$
transformation:
\begin{eqnarray}
M &\rightarrow& e^{2i\nu}\, M,  \nonumber\\
M' &\rightarrow& e^{-4i\nu}\, M'.
\end{eqnarray}
In order for the model to be consistent it is useful to introduce terms that mock up the axial and scale anomalies which read:
\begin{eqnarray}
&&\partial^{\mu}J^5_{\mu}=\frac{g^2}{16\pi^2}N_F\tilde{F}F=G
\nonumber\\
&&\theta^{\mu}_{\mu}=\partial^{\mu}D_{\mu} = -\frac{\beta(g)}{2g}FF=H.
\label{intr64553}
\end{eqnarray}
Note that in this section we use a different convention for the metric tensor with respect to section II.
Here $F$ is the $SU(3)_C$ field tensor, $\tilde{F}$ is its dual, $N_F$ is the number of flavors, $\beta(g)$ is the beta function for the coupling constant, $J^5_{\mu}$ is the axial current and $D_{\mu}$ is the dilatation current. The field $H$ may be associated in low energy QCD with the scalar glueball whereas $G$ with the pseudoscalar  one.
The Lagrangian  has the form
\begin{equation}
{\cal L} = - \frac{1}{2} {\rm Tr}
\left( \partial_\mu M \partial_\mu M^\dagger
\right) - \frac{1}{2} {\rm Tr}
\left( \partial_\mu M^\prime \partial_\mu M^{\prime \dagger} \right)
- V_0 \left( M, M^\prime \right) - V_{SB},
\label{mixingLsMLag}
\end{equation}
where $V_0(M,M^\prime)$ is a function made
from SU(3)$_{\rm L} \times$ SU(3)$_{\rm R}$
(but not necessarily U(1)$_{\rm A}$) invariants
formed out of $M$ and $M^\prime$.
The  leading choice of terms
corresponding
to eight or fewer underlying quark plus antiquark lines
at each effective vertex
reads \cite{Jora5}:
\begin{eqnarray}
V_0 =&-&c_2 \, {\rm Tr} (MM^{\dagger}) +
c_4^a \, {\rm Tr} (MM^{\dagger}MM^{\dagger})
\nonumber \\
&+& d_2 \,
{\rm Tr} (M^{\prime}M^{\prime\dagger})
+ e_3^a(\epsilon_{abc}\epsilon^{def}M^a_dM^b_eM'^c_f + {\rm H. c.})
\nonumber \\
&+&  c_3\left[ \gamma_1 {\rm ln} (\frac{{\rm det} M}{{\rm det}
	M^{\dagger}})
+(1-\gamma_1){\rm ln}\frac{{\rm Tr}(MM'^\dagger)}{{\rm
		Tr}(M'M^\dagger)}\right]^2.
\label{SpecLag}
\end{eqnarray}
All the terms except the last two (which mock up the axial anomaly)
have been chosen to also
possess the  U(1)$_{\rm A}$
invariance.

For a scalar and pseudoscalar Lagrangian of the type in Eq. (\ref{mixingLsMLag}) but that contains also scalar and pseudoscalar glubealls the trace anomaly reads \cite{Jora8}:
\begin{eqnarray}
\theta^{\mu}_{\mu}=M\frac{\partial V}{\partial M}+\frac{\partial V}{\partial M^{\dagger}}M^{\dagger}+4H\frac{\partial V}{\partial H}+4G\frac{\partial V}{\partial G}-4V,
\label{res553662}
\end{eqnarray}
where we applied the equation of motion in the results of section II.

From Eq. (\ref{finres663554}) we obtained in the previous section in terms of $\Phi$:
\begin{eqnarray}
&&\tau^{\mu}_{\mu}=\Bigg[-\frac{i}{3}\partial_{\mu}(\partial^{\mu}\Phi^*\Phi)+\frac{i}{3}\partial^{\rho}(\partial_{\rho}\Phi^*\Phi)\eta^{\mu}_{\mu}\Bigg]+h.c.=
\nonumber\\
&&-i\Bigg[\frac{\partial V}{\partial\Phi}\Phi-\frac{\partial V}{\partial \Phi^*}\Phi^*\Bigg].
\label{axanm657784}
\end{eqnarray}
Adjusted for the generalized linear sigma model the axial anomaly contribution in the regular case for a single chiral nonet $M$ is (Note that for this case the kinetic term has negative sign):
\begin{eqnarray}
\tau^{\mu}_{\mu}=i\Bigg[M\frac{\partial V}{\partial M}-M^{\dagger}\frac{\partial V}{\partial M^{\dagger}}\Bigg].
\label{res5346677}
\end{eqnarray}

Now we shall extend the scale transformation to include the axial one as discussed in section II. For that we denote $Q=H+iG$ and consider the generalized transformation:
\begin{eqnarray}
&&M\rightarrow \exp[-(1+i)\sigma]M(x\exp[-\sigma])
\nonumber\\
&&M^{\dagger}\rightarrow \exp[-(1-i)\sigma]M^{\dagger}(x\exp[-\sigma])
\nonumber\\
&&M'\rightarrow \exp[-(1-2i)\sigma[M'(x\exp[-\sigma])
\nonumber\\
&&M^{\prime\dagger}\rightarrow \exp[-(1+2i)\sigma[M^{\prime\dagger}(x\exp[-\sigma])
\nonumber\\
&&Q\rightarrow\exp[-4\sigma]Q(x\exp[-\sigma])
\nonumber\\
&&Q^{*}\rightarrow\exp[-4\sigma]Q^*(x\exp[-\sigma])
\label{res745536677}
\end{eqnarray}
The transformation in Eq. (\ref{res745536677}) makes sense as an expansion in the infinitesimal parameter $\sigma$.

Then the conservation of current should  be of the form:
\begin{eqnarray}
\theta^{\mu}_{\mu}+\tau^{\mu}_{\mu}=H-G,
\label{res664553}
\end{eqnarray}
where we used Eq. (\ref{intr64553}). Moreover we fixed the transformation law for the field $M'$ such that to correspond to the correct axial anomaly.  A detailed discussion of the most general structure of the possible terms that mock up the scale and trace anomalies is given in \cite{Jora8}. Here we will present a first order term that lead to the correct conservation law:
\begin{eqnarray}
&&V=V_1+V_1^{\dagger},
\label{defres666666}
\end{eqnarray}
where,
\begin{eqnarray}
&&V_1=Q\Bigg[\lambda_1\ln[\frac{Q}{\Lambda^4}]+\lambda_2\ln[\frac{\det M}{\Lambda^3}]+\lambda_3\ln[\frac{{\rm Tr}MM^{\prime\dagger}}{\Lambda^3}]\Bigg]+
\nonumber\\
&&V_1^{\dagger}=Q^*\Bigg[\ln[\lambda_1\frac{Q^*}{\Lambda^4}]+\lambda_2\ln[\frac{\det M^{\dagger}}{\Lambda^3}]+\lambda_3\ln[\frac{{\rm Tr}M'M^{\dagger}}{\Lambda^3}]\Bigg]
\nonumber\\
&&V=Q\Bigg[\lambda_1\ln[\frac{Q}{\Lambda^4}]+\lambda_2\ln[\frac{\det M}{\Lambda^3}]+\lambda_3\ln[\frac{{\rm Tr}MM^{\prime\dagger}}{\Lambda^3}]\Bigg]+
\nonumber\\
&&Q^*\Bigg[\ln[\lambda_1\frac{Q^*}{\Lambda^4}]+\lambda_2\ln[\frac{\det M^{\dagger}}{\Lambda^3}]+\lambda_3\ln[\frac{{\rm Tr}M'M^{\dagger}}{\Lambda^3}]\Bigg]
\label{lagrofinte6647783}
\end{eqnarray}

For the transformation introduced in Eq. (\ref{res745536677}) the trace of the scale and axial currents read:
\begin{eqnarray}
&&\theta^{\mu}_{\mu}=\Bigg[M\frac{\partial V}{\partial M}+M^{\dagger}\frac{\partial V}{\partial M^{\dagger}}+
M'\frac{\partial V}{\partial M'}+M^{\prime\dagger}\frac{\partial V}{\partial M^{\prime \dagger}}+4Q\frac{\partial V}{\partial Q}+4Q^{*}\frac{\partial V}{\partial Q^{*}}\Bigg]-4V
\nonumber\\
&&\tau^{\mu}_{\mu}=i\Bigg[M\frac{\partial V}{\partial M}-M^{\dagger}\frac{\partial V}{\partial M^{\dagger}}-
2M'\frac{\partial V}{\partial M'}+2M^{\prime\dagger}\frac{\partial V}{\partial M^{\prime \dagger}}\Bigg].
\label{res625535443}
\end{eqnarray}

We list below the contributions of various terms:
\begin{eqnarray}
&&M\frac{\partial V_1}{\partial M}=Q(3\lambda_2+\lambda_3)
\nonumber\\
&&\frac{\partial V_1^{\dagger}}{\partial M^{\dagger}}M^{\dagger}=Q^*(3\lambda_2+\lambda_3)
\nonumber\\
&&M'\frac{\partial V_1^{\dagger}}{\partial M'}=Q^*\lambda_3
\nonumber\\
&&\frac{\partial V_1}{\partial M^{\prime\dagger}}M^{\prime\dagger}=Q\lambda_3
\nonumber\\
&&4Q\frac{\partial V_1}{\partial Q}=4Q\Bigg[\lambda_1\ln[\frac{Q}{\Lambda^4}]+\lambda_2\ln[\frac{\det M}{\Lambda^2}]+\lambda_3\ln[\frac{{\rm Tr}MM^{\prime\dagger}}{\Lambda^2}]\Bigg]+4Q\lambda_1
\nonumber\\
&&4Q^*\frac{\partial V_1^{\dagger}}{\partial Q^*}=4Q^*\Bigg[\lambda_1\ln[\frac{Q^*}{\Lambda^4}]+\lambda_2\ln[\frac{\det M^{\dagger}}{\Lambda^2}]+\lambda_3\ln[\frac{{\rm Tr}M'M^{\dagger}}{\Lambda^2}]\Bigg]+4Q^*\lambda_1
\label{listofterms64553}
\end{eqnarray}
The final results are then for the combined trace and axial anomalies is:
\begin{eqnarray}
&&\theta^{\mu}_{\mu}=(H+iG)[3\lambda_2+2\lambda_3+4\lambda_1]+(H-iG)[3\lambda_2+3\lambda_3+4\lambda_1]=H[6\lambda_2+4\lambda_3+8\lambda_1]
\nonumber\\
&&\tau^{\mu}_{\mu}=i(H+iG)[3\lambda_2+3\lambda_3]-i(H-iG)[3\lambda_2+3\lambda_3]=-G[6\lambda_2+6\lambda_3].
\label{finalresults6477373}
\end{eqnarray}
Then a sufficient and necessary condition for Eq. (\ref{res664553}) to be fulfilled is:
\begin{eqnarray}
&&6\lambda_2+4\lambda_3+8\lambda_1=1
\nonumber\\
&&6\lambda_2+6\lambda_3=1
\label{finalcond66477}
\end{eqnarray}

As a consistency check it is important to notice that the trace and axial anomalies would be equally satisfied by the potential in Eq. (\ref{lagrofinte6647783}) also when the standard transformation for axial and trace anomalies would be applied.

\section{Proposal for a Lagrangian containing a complex glueball field}

In \cite{Jora8} an effective generalized linear sigma model that contained both scalar and pseudoscalar glueballs was introduced.  There $H$ and $G$ were introduced separately as individual states.  Here we will consider a new version of that Lagrangian where the scalar and pseudoscalar glueballs are parts of the same complex field $Q=H+iG$. This Lagrangian is consistent with the symmetries and axial and scale anomalies of low energy QCD and has the form:
\begin{eqnarray}
&&{\cal L}=-\frac{1}{2}{\rm Tr}[\partial^{\mu} M\partial_{\mu} M^{\dagger}]-\frac{1}{2}{\rm Tr}[\partial^{\mu}M' \partial_{\mu} M^{\prime\dagger}]-\frac{1}{2}[QQ^{*}]^{-\frac{3}{4}}[\partial^{\mu}Q\partial_{\mu}Q^*]+
\nonumber\\
&&f+f_{A,S}+V_{SB}.
\label{fulllagr54535}
\end{eqnarray}
In Eq. (\ref{fulllagr54535})  the kinetic term for the glueball field $Q$ is written such that to be scale invariant. The term $f$ can be:
\begin{eqnarray}
f =&-&c_2 \, {\rm Tr} (MM^{\dagger}) +
c_4^a \, {\rm Tr} (MM^{\dagger}MM^{\dagger})
\nonumber \\
&+& d_2 \,
{\rm Tr} (M^{\prime}M^{\prime\dagger})
+ e_3^a(\epsilon_{abc}\epsilon^{def}M^a_dM^b_eM'^c_f + {\rm H. c.})
\nonumber \\
&+& g_1\frac{QQ^*}{\Lambda^4},
\label{possibleter4657}
\end{eqnarray}
where $g_1$ is a dimensionless coupling constant and $\Lambda$ is the QCD scale.

Furthermore the axial and scale anomaly terms are encapsulated in $f_{A,S}=V$ where a detailed expression of $V$ is given in Eqs. (\ref{defres666666}) and (\ref{lagrofinte6647783}).  

The symmetry breaking terms can be:
\begin{eqnarray}
V_{SB}={\rm Tr}[A(M+M^{\dagger})],
\label{symnbre645}
\end{eqnarray}
where $A$ is a $3\times 3$ diagonal constant matrix proportional to the light quark masses.

Note that for each term in the full potential we chose the minimal ones from a multitude of possible terms.

\section{Conclusions}

In this work we explored some features of the scale transformation in the context of some simple theories in order to illustrate the possibility of incorporating both the scale and axial anomalies in a  single larger transformation. Our method is inspired by the supersymmetric theories where it is known that all anomalies including the trace and axial ones are part of the same supermultiplet structure \cite{Grisaru}. In order to express this at the level of currents we introduce a new tensor $\tau^{\mu}_{\nu}$ that has the correct properties such that the net anomalous conservation of current is given by the sum $\theta^{\mu}_{\mu}+\tau^{\mu}_{\nu}$. In some theories $\tau^{\mu}_{\mu}$ is a pseudoscalar.

This approach allowed us to introduce  a comprehensive term that mocks up both the trace and axial anomalies in an effective low energy QCD model. Consequently the scalar  $H$ and pseudoscalar $G$ glueballs that may be present in  a generalized linear sigma model with scalar and pseudoscalar meson nonets can be arranged in a  singlet complex field $H+iG$ on the same footing with the quark composite mesons. In this direction we also proposed a low energy effective Lagrangian based on the derivations and results in this work.  This method may have important application in low energy QCD models and besides that in any effective models that display both trace and axial anomalies.

\end{document}